**Initial incomplete author list, modify and add missing**
**Add your name and affiliation, author list is submitted separately to the abstract**

Mathieu Darnet, 1
Bitnarae Kim, 1
Simon Vedrine, 1
Jacques Deparis, 1
Francois Bretaudeau, 1
Julien Gance, 2
Fabrice Vermeersch, 2
Catherine Truffert, 2
Uula Autio, 3
Jochen Kamm, 3
Cedric Patzer, 3
Thomas Kalscheuer, 4
Suvi Heinonen, 5

Affiliations
1. BRGM, France
2. IRIS Instruments, France
3. Geological Survey of Finland, Finland
4. Uppsala university, Sweden
5. Institute of seismology, University of Helsinki, Finland

Short abstract:
In order to increase the mineral exploration success rate, the project SEEMS DEEP (SEismic and ElectroMagnetic methodS for DEEP mineral exploration) develops geophysical deep exploration workflow capable of imaging the bedrock from the surface down to several kilometres depth. In this paper, we present first results from ground electrical and electromagnetic surveys conducted at the SEEM DEEP geological test site, namely the Koillismaa Layered Intrusion Complex in north-eastern Finland. Here, a 1.7 km long hole drilled by GTK intersected mafic-ultramafic rocks with anomalous electrical and chargeability properties at ~1400 m depth, making it an interesting test site. To achieve this, we developed new sensors and survey protocols allowing to deploy in a cost-effective manner a large 3D grid of EM sensors (> 100 km$^2$). We also developed new protocols to deploy efficiently high-performance galvanic transmitters despite resistive grounds. Transmitting EM signals within such a large spread of live sensors allows to measure EM signals at long distances from the transmitters (> 10 km) and hence ensures a large depth of investigation (> 1 km). In addition, such an approach also allows to record IP signals associated to deep chargeable bodies.

# GROUND ELECTRICAL AND ELECTROMAGNETIC METHODS FOR DEEP MINERAL EXPLORATION – RESULTS FROM THE SEEMS DEEP PROJECT

**Introduction**

The transition towards carbon neutral transportation and energy sources increases the global demand for mineral raw materials while easy-to-find near-surface (< 200 m) ore deposits are unlikely discovered in well-explored areas such as Europe. In order to increase the mineral exploration success rate, the project SEEMS DEEP (SEismic and ElectroMagnetic methodS for DEEP mineral exploration) develops geophysical deep exploration workflow capable of imaging the bedrock from the surface down to several kilometres depth. In this paper, we present first results from ground electrical and electromagnetic surveys conducted at the SEEM DEEP geological test site, namely the Koillismaa Layered Intrusion Complex in north-eastern Finland. Here, a 1.7 km long hole drilled by GTK intersected mafic-ultramafic rocks with anomalous electrical and chargeability properties at ~1400 m depth, making it an interesting case study to test the ability of such technologies for imaging resistivity and chargeability contrasts at several kilometre depth.

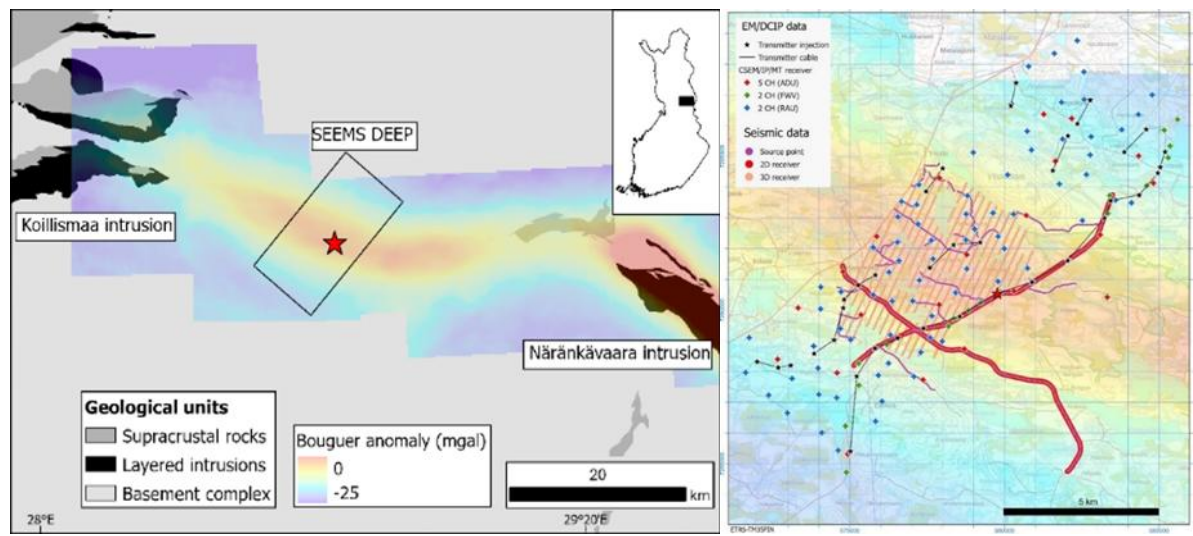

*Figure 1 (a) Koillismaa Layered Intrusion Complex consists of Koillismaa and Näränkävaara intrusions and the unexposed source of the connecting gravity anomaly. The geological test area for SEEMS DEEP (black rectangle) is centred on the GTK deep drill hole (red star). (b) Layouts of 2D and 3D seismic surveys and EM receivers and sources for combined CSEM, MT and ERT study.*

**Electromagnetic surveys**

SEEMS DEEP active electromagnetic surveys consisted in Controlled-Source-Electromagnetic (CSEM), Electrical Resistivity Tomography (ERT) and Induced Polarization (IP) surveys. In order to achieve a sufficient depth of investigation (> 1.5 km), a large 3D receiver and transmitter spread was deployed over a 200 km$^2$ area (Fig. 1). All receivers were autonomous and recording during the whole survey resulting in long-offset CSEM and ERT-IP signals (up to 13 km) that when modelled, should be capable of sensing deep conductivity and chargeability anomalies (> 2 km depth). Practically, we deployed 25 transmitter dipoles, each 1 km in length, and used three different galvanic transmitter systems (IRIS TIP 6000, Phoenix TXU-30 and Zonge GGT-3) capable of transmitting currents greater than a few Amperes in resistive grounds (> 1000 ohm-m), as found in such Archean basement rocks. Waveforms consisted in square waves with fundamental frequencies ranging from 0.125 Hz to 512 Hz and a 50% duty cycle wave at 0.0625 Hz. Transmitted signal durations were chosen to ensure a minimum of 100 stacks at all frequencies. On the receiver sides, we deployed 115 stations consisting in three types of systems: Metronix ADU08 MT recording electric and magnetic fields at 4096 Hz, IRIS Fullwaiver ERT recording electric fields at 100 Hz and BRGM made autonomous electric field stations based on Sercel RAU recording at 2000 Hz. In this paper, we report the results of the processing and inversion of the same dataset using both an ERT-IP and CSEM approach to assess the benefits and drawbacks of both approaches for deep resistivity and chargeability imaging.

### ERT-IP survey results

The ERT-IP survey data was obtained by extracting the DC response from all electric field stations. To do so, we processed the full time-series recorded at the four lowest main frequencies (from 0.0625 Hz to 8 Hz) to capture voltage data in a pseudo-steady state. Apparent resistivity data were obtained from the stacked voltage data. Traditionally, Induced Polarisation (IP) was computed in time domain (TDIP). Chargeability was initially defined as the ratio between the secondary voltage (voltage measured directly after the cancelling the current injection) and the primary voltage (before cancelling the current injection). In our case, the IP values were calculated in the frequency domain by measuring the delay between voltage and current (out-phasing). Note that some authors highlight correlation between out-phasing and computed chargeability in time domain (i.e. Binley, 2015). We applied a special processing routine to suppress Spontaneous Potential (SP) signals from the ERT-IP dataset. Firstly, we extracted a subset of the recorded data and removed the transient phase (i.e. front gate, when the current increase or decreases). Secondly, with statistical methods and using the alternative injections, the chargeability effect was estimated and removed from the signals. Thirdly, we removed the low-frequency trend from the data obtained in the second step by fitting the non-linear curve and assumed it removed the SP component. We inverted the ERT and IP data using the open-source code, pyGimLi (Rücker et al., 2017) in which the cost function is minimized using a Gauss-Newton scheme, to obtain a 3D resistivity and phase model. Filtering was performed by removing voltage values below 0.1 mV. Data with negative apparent resistivity is directly removed by pyGimLi. The inversion, which included a maximum of 4 iterations, was performed using 432 filtered data and unstructured mesh comprising 45,162 cells, applying basic roughness constraints.

The recovered resistivity model (Fig. 2) distinctly delineates a resistive layer at shallow depth and more conductive structures at deeper levels (> 1 km). The depth of investigation reaches approximately 2.5 km. The inverted phase model reveals a phase anomaly structure with a relatively high magnitude (about -25 mrad) starting below a depth of 1000 m, whereas IP values are nearly non-existent near surface. When comparing the 3D ERT and phase model with the existing geological model of the area, we observe a good correlation with the moderately conductive and chargeable body at 2 km depth and the ultramafic intrusion also observed in the drill core.

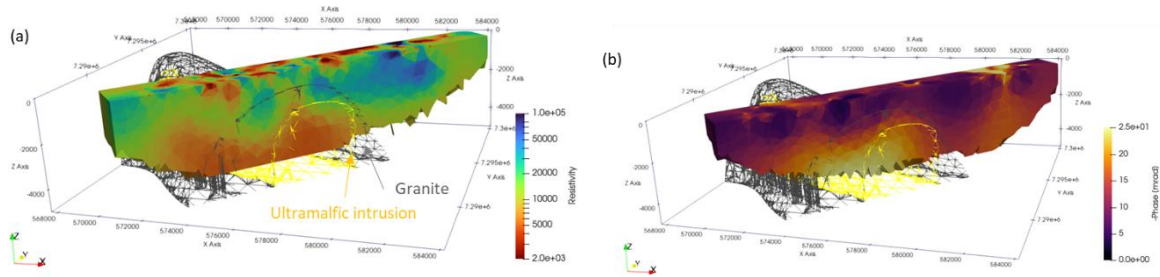

*Figure 2:* Cross section of the recovered (a) 3D resistivity model and (a) 3D phase model from the ERT-IP data. Grey mesh and yellow mesh represent granite and ultramafic intrusion models, respectively**.**

**CSEM survey results**

The CSEM data was obtained from extracting the Frequency-Domain Electro-Magnetic (FDEM) response from all electric field stations. To do so, we segmented all transmitter time series based on the start and stop times of each fundamental frequency of the square waves: 0.0625, 0.125, 0.5, 2, 8, 32, 128, and 512 Hz. Subsequently, for each receiver time series recorded within these intervals, we applied the frequency domain least-squares estimation algorithm implemented in the open-source code Razorback (Smaï and Wawrzyniak, 2020), to calculate the electric transfer functions and their error estimates at the fundamental frequency and 100 odd harmonics. A quality check was conducted to detect any outliers, employing a range of tools such as amplitude and phase spectra coherence, Argand diagrams, 1D inversions, and maps of the electric field vector orientations at receiver locations. Finally, data were computed using the semi-analytical code DIPOLE1D (Constable et al., 1987) within a synthetic 1D model derived from nearby resistivity core values. This synthetic data was compared to the measured field data from all transmitters which allowed us to check the consistency of the measured field data in terms of amplitudes and vector orientations with a 1D model response, and to detect any reverse polarity in the transmitter signals or instrumentation errors made during receiver deployment.

The QCed data was inverted in 3D using the open-source code custEM (Rochlitz et al., 2023), which relies on the inversion module pyGimLi and employs the Gauss-Newton (GN) inversion scheme with a Tikhonov regularization using a fixed roughness parameter. Two inversion runs were conducted: 1) a mono-frequency CSEM inversion at 0.125 Hz (1freq) and 2) a multi-frequency CSEM inversion at 0.125, 0.5, 2, 8, 32, 128, and 512 Hz (7freqs), covering the entire spectrum. We inverted both real and imaginary parts of the two horizontal components of the electric field using weights based on the inverse of standard errors estimated by the Razorback algorithm. Starting from the same homogenous 3D model of 10,000 ohm-m, the inversion '1freq' converged after 6 GN iterations from a root mean square (rms) value of 19.15 to 1.62, while inversion '7freqs' converged after 7 iterations from a rms value of 17.54 to 1.73. Both resulting resistivity models (Fig. 3) equally fit the 0.125 Hz frequency, yet display distinct structural geometries across the entire depth range. In order to assess the reliability of the '7freqs' model, we conducted a basic sensitivity test on deep structures by replacing all resistivity parameters below 2 km of depth in the last inverted model by the initial background value of 10,000 ohm-m, and then ran a new forward modelling. The previous rms value of 1.73 increased to 4.51 because of this change in resistivity. This indicates that resistivity variations at more than 2 km depth are constrained by the dataset. Similarly to the ERT results, when comparing the results with the existing geological model of the area, we observe a good correlation with the moderately conductive body at 2 km depth and the ultramafic intrusion.

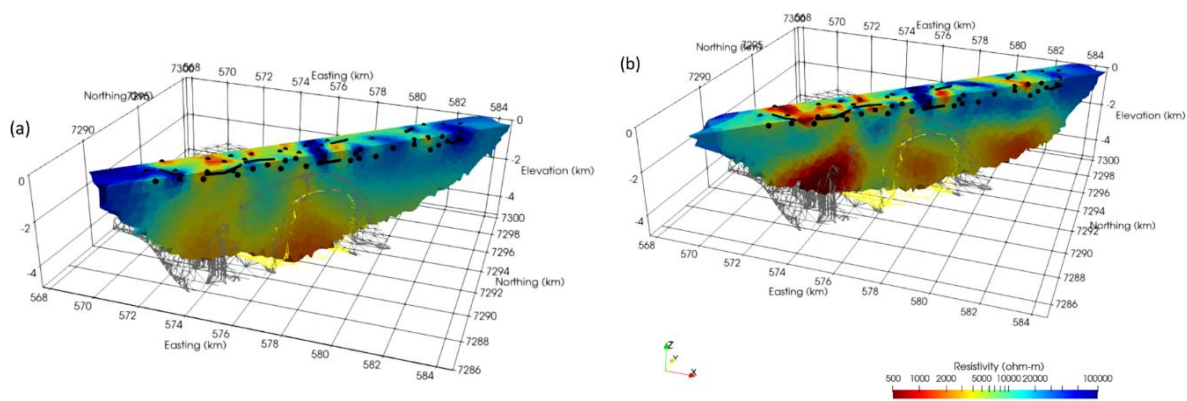

*Figure 3:* Cross section of the recovered (a) CSEM '1freq' resistivity model and (b) CSEM '7freqs' resistivity model. Grey mesh and yellow mesh represent granite and ultramafic intrusion models, respectively**.**

**Discussion**

The prime objective of this work was to test the ability of galvanically-coupled electromagnetic systems to image deep resistivity variations (> 1 km) in resistive environments as typically found in mineral exploration areas. To achieve this, we developed new sensors and survey protocols allowing to deploy in a cost-effective manner a large 3D grid of EM sensors (> 100 km$^2$). We also developed new protocols to deploy efficiently high-performance galvanic transmitters despite the resistive grounds (several amperes over kilometre long dipoles). Transmitting EM signals within such a large spread of live sensors allows to measure EM signals at long distances from the transmitters (> 10 km) and hence ensures a great depth of investigation (> 1 km), as demonstrated by the ERT and CSEM inversion results. In addition, such an approach also allows to record IP signals associated to deep chargeable bodies.

For this project, equipment, logistical and financial constraints limited the receiver spacing to 500 m minimum. Although this spacing is sufficient to detect deep resistivity and IP anomalies, it is insufficient to accurately map shallow resistivity and IP variations, which, in turn, will deteriorate the resolution at depth. To compensate for that, the multiple frequency approach showed promising results, as evidenced by the increased vertical resolution of the CSEM inversion results with multiple frequencies against one (Fig. 3a and b). Indeed, the higher the frequency of an EM signal, the shallower the skin depth is. Therefore, using a broad band of frequencies allows to discriminate better shallow anomalies (illuminated by both low and high frequencies) from deep anomalies (only illuminated by low-frequencies).

**Conclusions**

The SEEMS DEEP pilot survey conducted in the Koillismaa layered intrusion complex in Finland shows that the combination of careful survey design and technological innovations on EM sensors, transmitters and processing algorithms can significantly expand the operating envelop of existing technologies, especially towards underexplored areas at great depth (>1 km). Thanks to its ability to map conductors and chargeable targets, active electromagnetic surveying has a key role to play in the future exploration of deep mineral resources.

**Acknowledgements**

SEEMS DEEP project has been funded through ERA-MIN3 Joint Call 2021. National funding agencies: Finland: Business Finland (640/31/2022), France: ANR (ANR-22-MIN3-0006-02), Sweden: VINNOVA, Poland: NCBR (ERA-MIN3/1/113/SEEMSDEEP/2022) The SEEMS DEEP field crew is thanked for their tremendous effort in acquiring the seismic and EM data sets. Seismic receivers by FLEX-EPOS seismic instrument pool (Research Council of Finland 328776) and electric field sensors from the Geophysical Instrument Pool Potsdam (GIPP) were used in SEEMS DEEP measurements. Mira Geoscience is acknowledged for special licensing terms for Analyst Pro Geophysics software.